\documentclass{ws-ijmpe}

\begin{document}


\catchline{}{}{}{}{}

\title{Double-$\Lambda$ hypernuclei within a Skyrme-Hartree-Fock approach }

\author{\footnotesize Neelam Guleria$^{1}$, Shashi K. Dhiman$^{1,2}$, Radhey Shyam$^{3}$}

\address{$^1$Department of Physics, H. P. University, Shimla 171005, India \\
$^2$University Institute of Natural Sciences and Interface Technologies,
Himachal Pradesh Technical University, Hamirpur 177001, India \\
$^3$ Theory Division, Saha Institute of Nuclear Physics, 1/AF Bidhan Nagar,
Kolkata - 700064, INDIA and Physics Department, Indian Institute of Technology,
Roorkee, India}
\maketitle

\begin{history}
\received{received date}
\revised{revised date}
\end{history}
\begin{abstract}
We extend our Skyrme-Hartree-Fock (SHF) approach, which was used earlier to describe
successfully the single-$\Lambda$ hypernuclei, to investigate the binding and the
$\Lambda \Lambda$ bond energies of the double-$\Lambda$ hypernuclei. The 
nucleon-nucleon ($NN$) and $\Lambda$-nucleon ($\Lambda N$) interactions are taken 
from our previous study. For the $\Lambda \Lambda$ force several Skyrme-like 
potentials available in the literature have been used.  We discuss the sensitivity 
of the calculated $\Lambda \Lambda$ binding and bond energies to the  
$\Lambda \Lambda$ and $\Lambda N$ force parameters. It is found that the existing 
$\Lambda \Lambda$ hypernuclear data do not allow to distinguish between various 
$\Lambda \Lambda$ force parameter sets used by us. However, they show some 
selectivity for a particular set of the $\Lambda N$ potential determined in our 
previous work.

\end{abstract}


\section{Introduction}

A proper understanding of the properties of double-$\Lambda$ hypernuclei is important 
due to several reasons. These systems provide the unique opportunity to obtain 
information about the hyperon-hyperon ($YY$) interaction\cite{dal89}, which is crucial 
for  a complete understanding of the octet of baryons ($N, \Lambda, \Sigma, \Xi$) in 
a unified way. They also supplement the information about the hyperon-nucleon ($YN$) 
interaction that is mostly extracted from the studies of the single-$\Lambda$ 
hypernuclei\cite{has06}. The knowledge of $YN$ and $YY$ interactions is necessary 
for making extrapolations to understand the properties of both finite as well as bulk 
strange hadronic matter\cite{sch93} and the neutron stars\cite{sch08}. The study of 
the $\Lambda \Lambda$ hypernuclei is also of interest in connection with the possible 
existence of the strangeness ($S$) -2, six-quark $H$ dibaryon resonance with spin 
parity of $0^+$ and isospin 
$0$\cite{jaf77,mul83,bea11,ino11,sha11,hai11,hai12,bea11a,ino12,hai13,shy13,car13}.
Precisely measured binding energies of double-$\Lambda$ hypernuclei put a lower limit 
on the $H$ dibaryon mass\cite{ahn00,tak01,yoo07}. 

The double-$\Lambda$ hypernuclei were first observed in the 1960s\cite{dan63,pro66} 
in the studies of stopped $\Xi^-$ hyperons in emulsions. Two decades later, modern 
emulsion-counter hybrid techniques have been applied in the KEK-E176 experiment where a 
new double-$\Lambda$ hypernucleus event was found\cite{aok90,aok91}. Later on, in 
another hybrid emulsion experiment (KEK-E373) an unambiguous identification of the 
hypernucleus, ${_{\Lambda \Lambda}^6}$He, was made with a precise value of the 
binding energy of two $\Lambda$ hyperons\cite{tak01}. This is known as the NAGARA 
event.  Recently, in a reanalysis of the double-$\Lambda$ hypernuclear data produced 
in the KEK-E176 and KEK-E373 experiments, results for the $\Lambda \Lambda$ binding 
energies have been reported for ${_{\Lambda \Lambda}^6}$He, 
${_{\Lambda \Lambda}^{10}}$Be, ${_{\Lambda \Lambda}^{12}}$Be, and 
${_{\Lambda \Lambda}^{13}}$B hypernuclei\cite{nak10}.

These observations have led to a number of theoretical studies where several approaches 
have been used to investigate the double-$\Lambda$ hypernuclei. Calculations have 
been performed within the three- and four-body cluster models using the effective 
interactions or the 
$G$-matrices\cite{dal64,ali67,ban82,ike85,bod87a,bod87b,yam91,yam92,him93,hiy02}. 
Among other approaches are the Faddeev\cite{fil02}, variational 
Monte-Carlo\cite{usm04}, and variational six-body\cite{nem05} calculations. 
Furthermore, both nonrelativistic and relativistic mean field (RMF) models have also 
been used to predict the binding energies of such 
nuclei\cite{lan97,lan98,ruf87,ruf90,mar89,mar93,sch94,mar98,she06,hu13}. In 
Ref.~48 the two-$\Lambda$ binding energies of several double-$\Lambda$ nuclei 
between ${_{\Lambda \Lambda}^6}$He to ${_{\Lambda \Lambda}^{13}}$B were calculated 
within a shell model approach. Furthermore, using the interaction NSC97c of the 
Nijmegen group, the calculations for the $\Lambda \Lambda$ bond energies were reported
in Ref.\cite{vid04} within a $G$-matrix approach where the couplings between 
$\Lambda \Lambda$, $\Xi N$ and $\Sigma \Sigma$ channels were included. In these 
calculations moderate to good success has been achieved in predicting the 
two-$\Lambda$ binding and the $\Lambda \Lambda$ bond energies. 

The Skyrme-Hartree-Fock (SHF) model provides a self-consistent description of 
nuclear ground state properties\cite{vau72} and it has been shown to be a powerful 
tool  for investigating the gross properties of the nonstrange nuclei(see, e.g., 
a recent review\cite{erl11}). A clear advantage of this method is that it involves 
the complete summation of tadpole diagrams\cite{sai94,miy12}. The extension of 
this model to describe the single-$\Lambda$ hypernuclei was presented in 
Refs.~54 and 55. For calculating such strange nuclei, one requires, in addition to 
the Skyrme $NN$ force, also the Skyrme $\Lambda N$ interaction. In Ref.~56 the 
latter was determined from a Bruckner-Hartree-Fock calculation of the hypernuclear 
matter using the Nijmegen potentials NSC97a and NSC97f, which was used in an 
extended SHF scheme to determine the properties of single-$\Lambda$ hypernuclei. 
In this study the binding energies of the hypernuclei were somewhat overpredicted. 
In Ref.~57 several sets of the Skyrme $\Lambda N$ interactions were determined by 
fitting to the modern data on the binding energies of nearly twenty 
single-$\Lambda$ hypernuclei, which were used in the SHF model to describe the 
known properties of such nuclei over a wide mass range. 

In this paper, we present an extention of the SHF method of Ref.~57 to 
the calculations of the two-$\Lambda$ binding energies $(B_{\Lambda \Lambda})$ 
and the $\Lambda \Lambda$ bond energies $(\Delta B_{\Lambda \Lambda})$ of the 
double-$\Lambda$ hypernuclei. Unlike a few previous studies within similar approach
where calculations were limited to a few lighter systems, we have applied this
method to investigate the properties of the double-$\Lambda$ hypernuclei with 
masses covering essentially the entire range of the periodic table. In fact the 
SHF method provides an ideal approach for describing the heavier systems. 

In calculations of the double-$\Lambda$ hypernuclei, one needs as input the 
$\Lambda \Lambda$ interaction in addition to the $NN$ and $\Lambda N$ forces. 
Several phenomenological, meson-exchange motivated, and quark model based forms have 
been employed for the $\Lambda \Lambda$ force in the literature\cite
{bod87a,yam91,hiy02,sch94,vid04,car99,hai92,hai92a,alb02,afn03,fer05,fuj07,sch13}. 
We have, however, taken phenomenological Skyrme type of parameterizations for this 
force proposed in Refs.~39 and 66. In Ref.~39 three sets of parameters for the 
$\Lambda \Lambda$ interaction were determined by fitting to the $\Lambda \Lambda$ 
bond energy of $^{13}_{\Lambda \Lambda}$B ground state ($= 4.8 \pm 0.7$ MeV). 
In Ref.~66, four additional sets of the parameters were obtained by considering the 
results of a recent experimental analysis\cite{aok09}, in which a smaller bond 
energy ($=0.6 \pm 0.8$ MeV) for the $_{\Lambda \Lambda}^{13}$B ground state has 
been reported.  

We remark, however, that these forces are too simple and lack several important 
effects.  For example, they neglect the three-body interactions and the possible 
conversion of the $\Lambda \Lambda$ to $\Xi N$ and $\Sigma \Sigma$ channels. These 
constitute the important parts of the hyperon-hyperon interaction\cite{vid01,hai13} 
and neglecting them could bring in significant uncertainty in the calculations. 
Nevertheless, our aim in this paper is to extend and establish our SHF method for 
describing the double-$\Lambda$ hypernuclei. For this purpose we have taken these 
forces, which were also used in previous SHF type calculations of 
double-$\Lambda$ hypernuclei reported in Refs.~39, 66 and 68. Particularly 
noteworthy are the latter two references where these forces were employed to 
investigate the fission barriers of double-$\Lambda$ hypernuclei in the actinide 
region and the properties of neutron stars, respectively.  
  
\section{Formalism}

The total energy density functional (EDF) of a double-$\Lambda$ hypernucleus 
(${\cal E}^H_{2\Lambda}$) includes contributions from the total energy densities of 
nucleons (neutron and proton) (${\cal E}_{N}$) and of hyperons (${\cal E}_{\Lambda}$).
In addition, ${\cal E}^H_{2\Lambda}$ has terms arising from the pairing energy and 
the center of mass corrections, which are taken to be similar to those described in 
Ref.~57. ${\cal E}_{N}$ is related to the nucleon Hamiltonian density 
($H_N$) as
\begin{eqnarray}
{\cal E}_{N}& =& \int d^3r H_{N}({\bf r}).
\end{eqnarray}
The form of $H_N$ is the same as that given in Ref.\cite{gul12}. ${\cal E}_{\Lambda}$ 
is given by
\begin{eqnarray} \label{eq:ELambda}
{\cal E}_{\Lambda} & = & \int d^3r H_{\Lambda}({\bf r}).
\end{eqnarray}
In Eq.~(2) the hyperon Hamiltonian density, $H_\Lambda$, is the sum of two terms,
\begin{eqnarray}
H_{\Lambda}(\bf r) & = & H_{\Lambda N}({\bf r})+H_{\Lambda\Lambda}({\bf r}).
\end{eqnarray}
In Eq. (3), $H_{N\Lambda}({\bf r})$ has the same form as that described in 
Ref.~57 for the case of the single-$\Lambda$ hypernuclei. The second term, 
$H_{\Lambda \Lambda}$, is attributed to the $\Lambda \Lambda$ interaction and is 
given by
\begin{eqnarray} \label{H_LL}
H_{\Lambda\Lambda} & = & \frac{1}{4}\lambda_{0}\rho^{2}_{\Lambda}+
\frac{1}{8}(\lambda_{1}+3\lambda_{2})\rho_{\Lambda}\tau_{\Lambda}
 + \frac{3}{32}(\lambda_{2}-\lambda_{1})\rho_{\Lambda}\nabla^{2} \rho_{\Lambda}
\nonumber \\ 
& + & \frac{1}{4}\lambda_{3}\rho^{2}_{\Lambda}\rho^{\alpha}_{N},
\end{eqnarray}
where $\rho_{\Lambda}$ is the hyperon density and $\tau_{\Lambda}$ is the 
corresponding kinetic energy density. $\rho_N = \rho_p + \rho_n$, is the nucleon 
density. $\lambda_0$, $\lambda_1$, $\lambda_2$, and $\lambda_3$ are the parameters 
of the $\Lambda \Lambda$ force that will be discussed in the next section. The 
parameter $\alpha$ in the last term is assumed to be 1/3. In Eq.~(\ref{H_LL}), 
we have omitted terms corresponding to the $\Lambda$ spin density.

The wave functions for the proton, neutron and the $\Lambda$ particle are calculated 
from the SHF equations;
\begin{eqnarray}
\Bigg(-\frac{\hbar^{2}}{2m^{*}_{q}}{\bf \nabla}^{2}+V_{NN}({\bf r})+
V_q^{\Lambda}({\bf r})\Bigg)\phi_{q}({\bf r}) & = & \epsilon_{q}\phi_{q}
({\bf r}),
\end{eqnarray}
\begin{eqnarray}
\Bigg(-\frac{\hbar^{2}}{2m^{*}_{\Lambda}}{\bf \nabla}^{2}+V_\Lambda^{\Lambda }
({\bf r}) +V_{\Lambda\Lambda}({\bf r})\Bigg)\phi_{\Lambda}({\bf r}) & = & 
\epsilon_{\Lambda}\phi_{\Lambda}({\bf r}),
\end{eqnarray}
where $q$ represents a nucleon (proton or neutron), and $\epsilon_{q}$ and 
$\epsilon_{\Lambda}$ are the single-particle energies of the nucleon and the $\Lambda$ 
particle, respectively. The purely nuclear mean field potential [$V_{NN}({\bf r})]$, 
the additional field created by the $\Lambda$ hyperon that is seen by a nucleon 
[$V_q^{\Lambda}({\bf r})]$, and the whole nuclear field experience by a $\Lambda$ 
hyperon [$V_\Lambda^{\Lambda}({\bf r})$] have the same forms as those given in 
Ref.~57. $V_{\Lambda\Lambda}({\bf r})$, which is the field generated by the 
$\Lambda \Lambda$ interaction, is given by

\begin{eqnarray}
\label{pot_ll}
V_{\Lambda\Lambda}& = &\frac{1}{2} \lambda_{0}\rho_\Lambda+\frac{1}{8}
(\lambda_{1}+3\lambda_2)\tau_\Lambda + \frac{3}{16}(\lambda_2-\lambda_1)
\nabla^2\rho_\Lambda \nonumber \\ 
& + & \frac{1}{2}\lambda_3\rho_\Lambda \rho_N^\alpha.
\end{eqnarray}
The last term in Eq.~(\ref{pot_ll}) corresponds to the three-body $\Lambda\Lambda N$ 
interaction. In actual calculations, this term is dropped.  

While the nucleon effective mass remains the same as that described in 
Ref.~57, the $\Lambda$ effective mass, $m_\Lambda^*$, acquires additional 
terms due to the presence of $V_{\Lambda \Lambda}$,
\begin{eqnarray}
\frac{\hbar^{2}}{2m^{*}_{\Lambda}} & = & \frac{\hbar^{2}}{2m_{\Lambda}}+
\frac{1}{4}[u_{1} + u_{2}]\rho_{N}+
\frac{1}{8}[\lambda_{1}+3\lambda_{2}]\rho_{\Lambda},
\end{eqnarray}
where $u_1$ and $u_2$ are the parameters of the $\Lambda N$ force as defined 
in Ref.~57. Without the last term this equation is the same as that given  
in Ref.~57 for the effective mass of the single-$\Lambda$ hyperon. 

The main quantity in $\Lambda \Lambda$ hypernuclei is the $\Lambda \Lambda$ bond 
energy, which is defined as
\begin{eqnarray}
\Delta B_{\Lambda\Lambda} & = &B_{\Lambda\Lambda}-2B_{\Lambda},
\end{eqnarray}
where $B_{\Lambda}$ is the separation energy of one $\Lambda$ hyperon from the 
$^{A-1}_{\Lambda}Z$ hypernucleus and $B_{\Lambda\Lambda}$ is that of two $\Lambda$ 
hyperons from the $^{A}_{\Lambda\Lambda}Z$ hypernucleus, respectively. The separation 
energies are evaluated by solving the appropriate Hartree-Fock equations and by using 
the energy relations given in Ref.~57. In terms of the total binding energies
($E$), the bond energy can be expressed as $\Delta B_{\Lambda \Lambda} = 
E(^{A}_{\Lambda \Lambda}Z) + E(^{A-2}Z) - 2E(^{A-1}_{\Lambda} Z)$. It is clear that 
possible uncertainties in the center of mass treatment are mostly canceled in the bond 
energy.

\section{Results and discussions}

\begin{figure*}
\begin{center}
\psfig{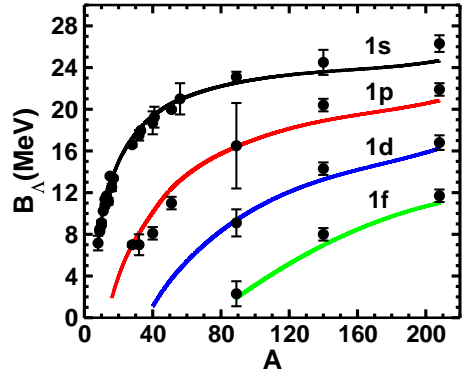} 
\end{center}
\caption{(Color online) Comparison of the $A$ dependence of the calculated
and experimental separation energies $B_\Lambda$ of $1s$, $1p$, $1d$ and $1f$ 
orbitals of the single-$\Lambda$ hypernuclei. The $NN$ and $\Lambda N$ interactions 
have been described by parameter sets SLy4 of Ref.\protect\cite{cha98} and 
HP$\Lambda$2 of Ref.\protect\cite{gul12}, respectively in each case. }
\end{figure*}

Before, presenting results for $B_{\Lambda \Lambda}$ and $\Delta B_{\Lambda \Lambda}$,
it would be of interest to discuss our SHF calculations for the single-$\Lambda$ 
separation energies $B_\Lambda$, as they will be used in obtaining $\Delta B_{\Lambda 
\Lambda}$ [see, Eq.~(9)]. This will be done very briefly here because all the details 
are given in\cite{gul12}. In Fig.~1, we show the $A$ dependence of the calculated 
$B_\Lambda$ of $1s$, $1p$, $1d$ and $1f$ shells of various single-$\Lambda$ 
hypernuclei. The corresponding available experimental data are also shown in this 
figure. In these calculations forces SLy4 (Ref.~69) and HP$\Lambda$2 (Ref.~57) have 
been used for $NN$ and $\Lambda N$ interactions, respectively. We note that apart 
from the A=208 case, where we see some underbinding of the $1s$ orbital, our 
calculations are in close agreement with the  experimental $B_\Lambda$. Thus our 
SHF model with $\Lambda N$ force HP$\Lambda$2 provides a good description of the 
single-$\Lambda$ separation energies for the lighter as well as heavier systems.
\begin{table}
\begin{center}
\caption{Parameters for the  $\Lambda\Lambda$ interaction. The last column gives the 
ranges of "equivalent single Gaussian" potentials. 
}
\label{tab-para-ll}
\begin{tabular}{cccccc}
\hline
SET & $\lambda_0$ & $\lambda_1$ & $\mu$ \\
 & (MeV fm$^3$)& (MeV fm$^5$) &  (fm)\\
\hline
\hline
S$\Lambda\Lambda 1$        & -312.6 &  57.5 & 0.61 \\
S$\Lambda\Lambda 2$        & -437.7 & 240.7 & 1.05 \\
S$\Lambda\Lambda 3$        & -831.8 & 922.9 & 1.49  \\
S$\Lambda\Lambda 1^\prime$&  -37.9 &  14.1 & 0.61 \\
S$\Lambda\Lambda 3^\prime$& -156.4 & 347.2 & 1.49 \\
\hline
\end{tabular}
\end{center}
\end{table}

We now proceed to the discussion of the double-$\Lambda$ hypernuclei, which is 
the main focus of this paper. In calculations of the binding energies of the 
double-$\Lambda$ hypernuclei, we have used the parameter sets SLy4 and HP$\Lambda$2 
for the $NN$ and the $\Lambda N$ forces, respectively. For the $\Lambda \Lambda$ 
force, the three parameter sets (S$\Lambda \Lambda 1$, S$\Lambda \Lambda 2$ and 
S$\Lambda \Lambda 3$) are reported in Ref.\cite{lan98} (see Table~\ref{tab-para-ll}),
have been employed. In these sets the density dependent terms of Eq.~(4) 
($\lambda_2$ and $\lambda_3$) were ignored because the $p$ wave contributions do not 
take part in the lowest single-particle level. The parameters $\lambda_0$ and 
$\lambda_1$ were determined by fitting to the bond energy $\Delta B_{\Lambda \Lambda}
= 4.8 \pm 0.7$ MeV of the $^{13}_{\Lambda \Lambda}$B ground state. The difference 
between the three parameter sets is the interaction range $\mu = 
\sqrt{-\lambda_1/\lambda_0}$, which is evaluated from the equivalent single-Gaussian 
potential assumption. In addition, two  more sets
of the $\Lambda \Lambda$ force parameters were obtained in Ref.\cite{min11} 
(sets S$\Lambda \Lambda 1^\prime$ and S$\Lambda \Lambda 3^\prime$ of 
Table~\ref{tab-para-ll}) by fitting to a weaker bond energy $\Delta B_{\Lambda 
\Lambda} = 0.6 \pm 0.8$ for the $^{13}_{\Lambda \Lambda}$B ground state reported in 
Ref.~67. This results from taking into account the excited state of the 
daughter single-$\Lambda$ hypernucleus $_{\Lambda}^{13}$C$^*$ in the decay channel. 
We have used these 5 sets of the $\Lambda \Lambda$ forces in our study. The parameter
sets S$\Lambda \Lambda R1$ and S$\Lambda \Lambda R2$ of Ref.\cite{min11} that 
correspond to a repulsive $\Lambda \Lambda$ interaction, have also been used in few 
cases. However, these sets produce unrealistic bond energies in our calculations so
we do not discuss them here.

In Table~\ref{tableHPL2}, we show our results for the double-$\Lambda$ binding 
energies for a number of hypernuclei that are obtained by using sets 
S$\Lambda \Lambda 1$, S$\Lambda \Lambda 2$ and S$\Lambda \Lambda 3$ for the 
$\Lambda \Lambda$ force. In each case, the parameter sets SLy4, and HP$\Lambda$2 
were employed for $NN$ and $\Lambda N$ interactions, respectively. In this table
we have also listed the $B_\Lambda$ ($^{A-1}{_\Lambda Z}$),  obtained with  
the same $NN$ and $\Lambda N$ forces. It is seen that $B_{\Lambda \Lambda}$ 
increases with mass number of the hypernucleus. This is in agreement with the 
observations made in the RMF calculations of the double-$\Lambda$ hypernuclei in 
Refs.~45 and 46. We further note that for the first two lightest mass 
hypernuclei, $B_{\Lambda \Lambda}$ depends rather strongly on the $\Lambda \Lambda$ 
force. However, with increasing mass this dependence becomes less stronger. 

For the purpose of comparison we have also shown in Table~\ref{tableHPL2}, the 
single-$\Lambda$ separation energies $B_\Lambda$ for the same systems. We see
that for double-$\Lambda$ systems heavier than mass 10, the $B_{\Lambda \Lambda}$ 
is nearly twice of the $B_\Lambda$. However, for the lighter systems, this 
is not so. It is shown in Ref.~46 that $B_{\Lambda \Lambda}$ is related to the 
$\Lambda$ single-particle energies $(\epsilon_\lambda)$ as $B_{\Lambda \Lambda} 
= 2\epsilon_\lambda - E_R$,  where $E_R$ is the rearrangement energy that 
quantifies the core polarization. Therefore, results shown in 
Table~\ref{tableHPL2} indicate that the core polarization effects may contribute 
significantly to the binding energy for lighter systems. More discussion on the  
core polarization effect is presented towards the end of this section. 

For $^6_{\Lambda \Lambda}$He and $^{10}_{\Lambda \Lambda}$Be, the $B_{\Lambda 
\Lambda}$ calculated with force S$ \Lambda \Lambda 3$ that has the largest range, 
reproduce the corresponding experimental values the best. The other two sets lead 
to larger binding energies for these hypernuclei. This could indicate that a 
longer range $\Lambda \Lambda$ force leads to a lesser binding of two $\Lambda$s 
to a lighter core.  Nevertheless, it should be emphasized that description of the 
lightest nuclei $^6_{\Lambda \Lambda}$He and $^{10}_{\Lambda \Lambda}$Be may be less 
reliable in the SHF approach\cite{lan98,sch13}. The cluster\cite{hiy10,hiy10a,hiy12}
or the shell model\cite{gal11} methods should be more appropriate for these cases.  

\begin{table*}[t]
\begin{center}
\caption{B$_{\Lambda\Lambda}$ ($^A_{\Lambda \Lambda}Z)$ of various hypernuclei 
($^A_{\Lambda \Lambda}Z)$ calculated using sets SLy4,  
HP$\Lambda 2$ of Ref.\protect\cite{gul12}, for $NN$, and $\Lambda$N interactions, 
respectively, and the three parameter sets S$\Lambda \Lambda 1$, S$\Lambda \Lambda 2$, 
and S$\Lambda \Lambda 3$ for the $\Lambda \Lambda$ force. For comparison the 
separation energies [$B_\Lambda$($^{A-1}_{\Lambda}Z$)] of the single-$\Lambda$ 
hyperon calculated with forces SLy4, and HP$\Lambda 2$ are also shown.}
\label{tableHPL2}
\begin{tabular}{|c|c|cccc|c|}
\hline
Hypernuclei &$B_{\Lambda}$
& B$_{\Lambda\Lambda}^{S{\Lambda\Lambda1}}$
& B$_{\Lambda\Lambda}^{S{\Lambda\Lambda2}}$ 
& B$_{\Lambda\Lambda}^{S{\Lambda\Lambda3}}$ 
& B$_{\Lambda\Lambda}^{(Exp.)}$ &Ref., Event \\
($^A_{\Lambda \Lambda}Z)$ & {\footnotesize (MeV)}&{\footnotesize (MeV)} 
&{\footnotesize (MeV)} & {\footnotesize (MeV)} & {\footnotesize (MeV)}\\
\hline
$^{6}_{\Lambda\Lambda}$He &7.12& 11.88 & 9.25 & 7.60 & $6.91\pm 0.16$ &
Ref.\cite{nak10}, NAGARA\\
$^{10}_{\Lambda\Lambda}$Be&10.76 & 19.78 & 18.34 & 15.19 & $14.94\pm0.13^*$ &
Ref.\cite{dan63}\\
$^{11}_{\Lambda\Lambda}$Be&10.80 & 20.55 & 19.26 & 16.27 & $20.49\pm1.15$ &
Ref.\cite{nak10}, HIDA \\
$^{12}_{\Lambda\Lambda}$Be&10.91 & 21.10 & 19.97 & 17.18 & $22.23\pm1.15$ &
Ref.\cite{nak10} \\
$^{13}_{\Lambda\Lambda}$B&11.02 & 21.21 & 20.26 & 17.76 & $23.30\pm0.70$ &
Ref.\cite{nak10}, E176 \\
\hline
\end{tabular}
\end{center}
*This value has been obtained from the experimentally deduced value $11.90\pm0.13$
MeV by adding 3.04 MeV for the 2$^+$ excitation energy, assuming equal 2$^+$
core excitation energies in $^9_\Lambda$Be and 
$^{10}_{\Lambda\Lambda}$Be\cite{gal11}.
\end{table*}

On the other hand, for hypernuclei $^{11}_{\Lambda \Lambda}$Be, $^{12}_{\Lambda 
\Lambda}$Be, and $^{13}_{\Lambda \Lambda}$B, $B{_{\Lambda \Lambda}}$s calculated 
with sets S$\Lambda \Lambda 1$ and S$\Lambda \Lambda 2$ do not differ much from 
each other and reproduce the experimental data better in comparison to those 
obtained with set S${\Lambda \Lambda 3}$. The SHF method is expected to be 
relatively better suited to describe these systems. 

It would be interesting to compare our $B_{\Lambda \Lambda}$ with the available 
corresponding results obtained in other theoretical approaches. For the nucleus 
$^{11}_{\Lambda \Lambda}$Be, results for the binding energy are available in both 
the shell model as well as the cluster model methods. We note that while both 
$B_{\Lambda \Lambda}^{S\Lambda \Lambda 1}$ and 
$B_{\Lambda \Lambda}^{S\Lambda \Lambda 2}$ of 
this hypernucleus are larger than that the value (18.40 MeV) predicted by the shell 
model calculation of Ref.\cite{gal11} by about 5-10$\%$, they are comparable to that 
calculated (19.81 MeV) in a $\alpha \alpha n \Lambda \Lambda$ five-body cluster 
model\cite{hiy10}. Furthermore, our $B_{\Lambda \Lambda}$s are similar to that 
obtained (19.46 MeV) recently within a quark mean-field model\cite{hu13}. For the 
$^{12}_{\Lambda \Lambda}$Be and $^{13}_{\Lambda \Lambda}$B hypernuclei, our binding 
energies are about 10-15$\%$ larger than the values predicted by both the shell 
model and the quark mean-field model. It is worth noticing that our results are in 
agreement with the corresponding experimental data within the statistical error 
except for the $^{13}_{\Lambda \Lambda}$B case where our calculations underpredict 
the data.

In Table 3, we display our results for the bond energy $\Delta B_{\Lambda 
\Lambda}$ calculated with $\Lambda \Lambda$ forces S$\Lambda\Lambda$1, 
S$\Lambda\Lambda$2 and S$\Lambda\Lambda$3. These are obtained by using Eq.~(9), 
where the $B_\Lambda$ values have been taken from the Ref.~57 (also shown 
in Fig.~1). The experimental points are from Refs.\cite{nak10,aok09}. It is seen 
that generally $\Delta B_{\Lambda \Lambda}$ decreases with increasing $A$. It is 
further noted that for more complex (heavier) double-$\Lambda$ hypernuclei the 
bond energies are smaller. This points to the fact that for heavier systems the 
binding energies $B_{\Lambda \Lambda}$ are closer to the twice of $B_\Lambda$, which 
is seen already in Table 2. These results are similar to those obtained in the SHF 
and RMF calculations of Refs.~39 and 45, respectively. 
\begin{table}[h]
\begin{center}
\caption{Bond energies $\Delta B_{\Lambda \Lambda}$ calculated with 
$\Lambda \Lambda$ interactions S$\Lambda\Lambda$1, S$\Lambda\Lambda$2 and 
S$\Lambda\Lambda$3. In each case the parameter sets SLy4 and HP$\Lambda$2 were
employed for the $NN$ and $\Lambda N$ interactions, respectively. The total baryon 
number of the double-$\Lambda$ hypernucleus is represented by $A$ in the first 
column.}
\label{bond-energy}
\vskip 0.3cm
\begin{tabular}{c|ccc|c}
\hline
$A$ && $\Delta B_{\Lambda\Lambda}$ && $\Delta B_{\Lambda\Lambda}$(exp.) \\
    & S$\Lambda\Lambda$1 & S$\Lambda\Lambda$2 & S$\Lambda\Lambda$3  & \\
    & {\footnotesize (MeV)} & {\footnotesize (MeV)} & {\footnotesize (MeV)} & 
 {\footnotesize (MeV)}\\
\hline
06 & 2.36 & 4.99 & 6.64 & $3.82\pm1.72$\\
09 & 2.17 & 4.71 & 5.5 & \\
10 & 1.72 & 3.64 & 4.27 & $1.3\pm0.4$ \\
11 & 0.90 & 2.74 & 3.91 & $2.27\pm1.23$ \\
12 & 0.71 & 1.97 & 2.44 & \\
13 & 0.69 & 1.71 & 2.32 & $0.6\pm0.8$ \\
30 & 0.55 & 1.33 & 1.56 &\\
50 & 0.50 & 1.01 & 1.11 &\\
58 & 0.45 & 0.99 & 1.08 &\\
92 & 0.33 & 0.73 & 0.91 &\\
140 & 0.31 & 0.58 & 0.77 &\\
210 & 0.25 & 0.42 & 0.48 &\\
\hline
\end{tabular}
\end{center}
\end{table}
   
In Fig.~2(a) the $A$ dependence  of $\Delta B_{\Lambda \Lambda}$ is shown in some 
more details. We see that whereas the decrease of $\Delta B_{\Lambda \Lambda}$ with 
increasing $A$ is quite steep for lighter hypernuclei, it is gradual for the medium 
mass and heavier systems. We further note that with the S$\Lambda \Lambda1$ force, 
the agreement between the calculated and the experimental bond energy for the last two 
data points is somewhat better as compared to that obtained with sets S$\Lambda 
\Lambda 2$ and S$\Lambda \Lambda 3$ - the bond energy determined with set 
S$\Lambda \Lambda 3$ is farthest from the data for these points. However, given 
the large statistical errors in the data points it is premature to draw any definite 
conclusion about the preference of one parameter set over the other. More 
experimental data, particularly for heavier double-$\Lambda$ hypernuclear systems 
are needed to extract an unambiguous information about the $\Lambda \Lambda$ force 
from such calculations.  

In fig.~2(b), we present a comparison of the $A$ dependence of $\Delta B_{\Lambda 
\Lambda}$ obtained with parameters sets S$\Lambda \Lambda 1^\prime$ and S$\Lambda 
\Lambda 3^\prime$ of Ref.\cite{min11}, and S$\Lambda \Lambda 1$. We see that 
set S$\Lambda \Lambda 3^\prime$ leads to the $\Delta B_{\Lambda \Lambda}$ 
that are larger in magnitude and fall less steeply with increasing A as compared to
those obtained with set S$\Lambda \Lambda 1$. On the other hand, bond energies 
produced by set S$\Lambda \Lambda 1^\prime$ are comparable with those of set 
S$\Lambda \Lambda 1$ for $A > 50$. However, for $A < 50$ the difference between the 
two is quite large. It should be remarked that the $\Delta B_{\Lambda \Lambda}$ 
calculated with parameter sets S$\Lambda \Lambda 3^\prime$ and 
S$\Lambda \Lambda 1^\prime$ for $A$ around 10 are larger than the value 0.6$\pm$0.8 
MeV, to which they are fitted to in Ref.~66. This can be understood from 
the fact that in the fitting procedure of Ref. 66, the adopted $NN$ and 
$\Lambda N$ interactions were different from those used in our calculations. As will 
be shown later on, the calculated $\Delta B_{\Lambda \Lambda}$ shows strong 
dependence over $\Lambda N$ interaction.  
\begin{figure*}
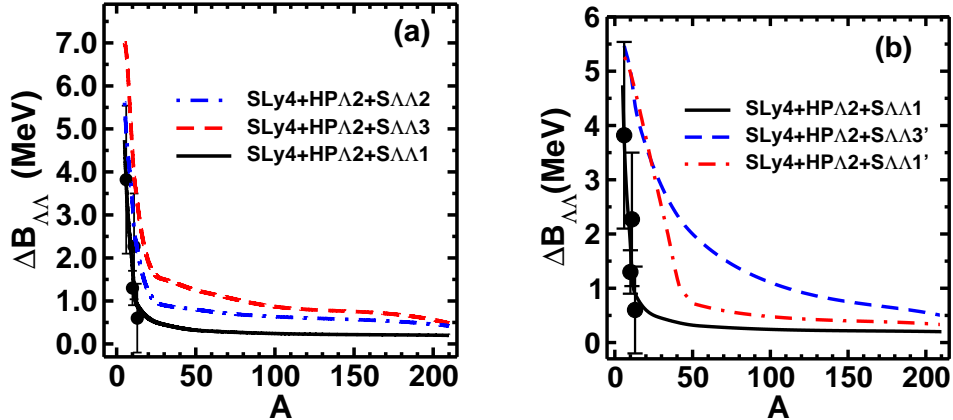

\begin{center}
\begin{tabular}{cc}
\psfig{file=Fig2a.eps,width=6cm} &\hspace{0.5cm}
\psfig{file=Fig2b.eps,width=5.5cm}
\end{tabular}
\end{center}
\caption{(Color online)(a) Comparison of the $A$ dependence of the bond 
energy $\Delta B_{\Lambda \Lambda}$ obtained by using  parameter set S$\Lambda \Lambda 
1$, S$\Lambda \Lambda 2$, and S$\Lambda \Lambda 3$ for the $\Lambda \Lambda$
force.  The $NN$ and $\Lambda N$ interactions have been described by parameter sets 
SLy4 and HP$\Lambda$2, respectively in each case. (b) Same as in (a) for $\Lambda 
\Lambda$ force parameter sets S$\Lambda \Lambda 1$, S$\Lambda \Lambda 1^\prime$, 
S$\Lambda \Lambda 3^\prime$. The experimental points are taken from 
Refs.\protect\cite{nak10,aok09} }
\end{figure*}

In Fig.~3, we show the sensitivity of $\Delta B_{\Lambda \Lambda}$ to the $\Lambda N$ 
force. In these calculations parameters sets SLy4 and S$\Lambda\Lambda 1$ have been 
used for the  $NN$ and $\Lambda \Lambda$ interactions, respectively. For the 
$\Lambda N$ force we have used sets HP$\Lambda 2$, N$\Lambda 1$ and O$\Lambda 1$ of 
Ref.\cite{gul12}.  It may be recalled here that while the set HP$\Lambda 2$ provides 
a good agreement with the experimental binding energies of the $\Lambda$ 
single-particle states of all the orbitals in the entire mass range of hypernuclei 
(see Fig.~1), the parameter sets O$\Lambda 1$ and N$\Lambda 1$ slightly overestimate  
the data for the lighter nuclei (we refer to Ref.~57 for a detailed discussion 
of this points).  In Fig.~3, we note that even the currently available sparse data 
clearly favor the parameter set HP$\Lambda 2$.   

It should, however, be added that the bond energy as defined by Eq.~(9) is more 
applicable to those cases where the core nuclei $^{A-1}_\Lambda$Z have zero spin. 
In case of the nonzero spin core nucleus, the $B_\Lambda$ appearing in Eq.~(9) is 
actually an average of the binding energies of the spin-doublet 
states\cite{lan98,hiy02}. The bond energy is also influenced by the structural 
changes that are caused to the core nucleus due to the $\Lambda$-core interaction 
(e.g., the core polarization\cite{lan98}). The core polarization effects are 
significant for $\Lambda N$ potentials that are strongly polarizing. However, the 
$\Lambda \Lambda$ forces used in this study have been extracted by fitting the data 
with SHF calculation where the used $\Lambda N$ interactions lead to small core 
polarization energy in $^{12}_\Lambda$B\cite{lan98}. Furthermore, the spin-doublet 
splitting in $^{12}_\Lambda$B is probably small\cite{yam94}. Nevertheless, an 
alternative definition of the $\Lambda \Lambda$ bond energy is suggested in 
Ref.~34, where it is essentially determined by the strength of the $\Lambda
\Lambda$ interaction. However, the prevailing uncertainties in this interaction may 
also creep into the bond energies calculated within this alternative method. 
\begin{figure*}
\begin{center}
\psfig{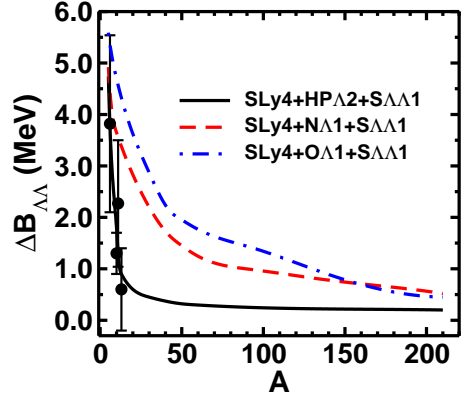}
\end{center}
\caption{(Color online)$A$ dependence of the bond energy
$\Delta B_{\Lambda \Lambda}$ for three parameter sets of $\Lambda N$
force. $NN$ and $\Lambda \Lambda$ interactions have been described by
parameter sets SLy4 and S$\Lambda\Lambda 1$, respectively, in each case.}
\end{figure*}

\section{Summary and conclusions}

In conclusion, we have calculated the binding energies and the bond energies of the 
light to heavy double-$\Lambda$ hypernuclei within a Skyrme-Hartree-Fock model. This 
is an extension of the model used earlier to describe successfully the binding 
energies of the $\Lambda$ single particle states of both lighter as well as heavier 
single-$\Lambda$ hypernuclei\cite{gul12}. For the $NN$ interaction the parameter set 
SLy4 of Ref.\cite{cha98} has been used while for the $\Lambda N$ force the parameter 
set HP$\Lambda$2 of Ref.\cite{gul12} has been employed. These sets provide a 
reasonable overall description of the single-$\Lambda$ hypernuclear data. For the 
$\Lambda \Lambda$ force, parameter sets S$\Lambda \Lambda 1$, S$\Lambda \Lambda 2$ 
and S$\Lambda \Lambda 3$ of Ref.\cite{lan98} as well as S$\Lambda \Lambda 1^\prime$ 
and S$\Lambda \Lambda 3^\prime$ of Ref.\cite{min11} were used. Since in this work our 
aim has been more to establish our SHF model for the description of the 
double-$\Lambda$ hypernuclei, we selected the $\Lambda \Lambda$ force parameters 
that are already available in the literature. In future, efforts will be made to 
determine corresponding force parameters by refitting the data using our 
$NN$ and $\Lambda N$ interactions. Nevertheless, even such a force will not be free 
from uncertainties.

We have calculated the binding energies of a number of double-$\Lambda$ hypernuclei
where some experimental information is available. We showed that SHF calculations 
done with empirical Skyrme type $\Lambda \Lambda$ forces without the density 
dependent terms, provide a reasonable description of the $\Lambda \Lambda$ 
hypernuclear systems. Our results for the two-$\Lambda$ binding energies of the 
hypernuclear systems $^{11}_{\Lambda \Lambda}$Be, $^{12}_{\Lambda \Lambda}$Be, and 
$^{13}_{\Lambda \Lambda}$B are comparable to those obtained within other approaches 
such as the shell model, the cluster model and the quark mean-field model.  
However, in the present calculation it has not been possible to reproduce 
simultaneously the experimental binding energies of all the known double-$\Lambda$ 
hypernuclei with any one set of the $\Lambda \Lambda$ potential. Further studies are
required to obtain more precise information about this force as compared to what is 
available now.
    
We have also studied the $A$ dependence of the $\Lambda \Lambda$ bond energy in 
the ground state of the double-$\Lambda$ hypernuclei. We observe that the currently
available limited experimental data for such hypernuclei do not allow to distinguish 
between the $\Lambda \Lambda$ forces used in this study.  However, they show a 
significant selectivity for the $\Lambda N$ force where the set HP$\Lambda$2 is 
favored. We acknowledge that the $\Lambda \Lambda$ forces used by us are too 
simple. Moreover, we have not considered the core polarization effects, which  
depend on the $\Lambda N$ interaction\cite{lan98}. However, calculations made with 
more realistic $\Lambda \Lambda$ potentials in Ref.\cite{lan97} arrive at similar 
conclusions. At the present stage of our knowledge on hyperon-nucleon and 
hyperon-hyperon interactions any fine tuning of these forces would clearly be 
premature. 

A systematic study of the data over a large mass range - as is done in the present 
paper, is necessary for deriving constraints on various interactions and density 
functionals. More experimental information on the double-$\Lambda$ hypernuclei over 
a wide mass range is, therefore, clearly required. The mean field method, on the other
hand, may come to its limit for very light nuclei, but experience with SHF 
calculations on nonstrange nuclei do not show dramatic failures of the method for
such systems. Rather they are surprisingly successful even in  mass-4 region. Thus
this method is quite robust. Our work demonstrates that the Skyrme-Hartree-Fock model
can be used as a workable theoretical framework for investigating the properties of 
both single- and double-$\Lambda$ hypernuclei over a wide mass region. 

\section{Acknowledgments}

One of the authors (NG) would like to thank the theory division of the Saha Institute 
of Nuclear Physics for the kind hospitality during her several visits there. This work 
has been supported by the Council of Scientific and Industrial Research (CSIR), India.

\end{document}